Crash Themes in Automated Vehicles: A Topic Modeling Analysis of the California Department of Motor Vehicles Automated Vehicle Crash Database

Hananeh Alambeigi
Texas A&M University Department of Industrial and Systems Engineering
Email: hana.alambeigi@tamu.edu
ORCID: 0000-0003-4310-3950

Anthony D. McDonald, Corresponding Author
Texas A&M University Department of Industrial and Systems Engineering
Email: mcdonald@tamu.edu
ORCID: 0000-0001-7827-8828

Srinivas R. Tankasala
Texas A&M University Department of Industrial and Systems Engineering
Email: tankr98@tamu.edu

**ABSTRACT**
Automated vehicle technology promises to reduce the societal impact of traffic crashes. Early investigations of this technology suggest that significant safety issues remain during control transfers between the automation and human drivers and automation interactions with the transportation system. In order to address these issues, it is critical to understand both the behavior of human drivers during these events and the environments where they occur. This article analyzes automated vehicle crash narratives from the California Department of Motor Vehicles automated vehicle crash database to identify safety concerns and gaps between crash types and current areas of focus in the current research. The database was analyzed using probabilistic topic modeling of open-ended crash narratives. Topic modeling analysis identified five themes in the database: driver-initiated transition crashes, sideswipe crashes during left-side overtakes, and rear-end collisions while the vehicle was stopped at an intersection, in a turn lane, and when the crash involved oncoming traffic. Many crashes represented by the driver-initiated transitions topic were also associated with the side-swipe collisions. A substantial portion of the side-swipe collisions also involved motorcycles. These findings highlight previously raised safety concerns with transitions of control and interactions between vehicles in automated mode and the transportation social network. In response to these findings, future empirical work should focus on driver-initiated transitions, overtakes, silent failures, complex traffic situations, and adverse driving environments. Beyond this future work, the topic modeling analysis method may be used as a tool to monitor emergent safety issues.

# 1 INTRODUCTION

Traffic crashes have a significant impact on the world economy and are a leading cause of death and injuries worldwide (*1*, *2*). Advanced safety technologies such as forward collision warnings, blind spot monitoring, and automated emergency braking have reduced crashes and crash severity (*3*–*6*). Automated vehicle technologies—such as the "Tesla Autopilot" (*7*), "IntelliSafe Autopilot" (*8*), "GM Cruise" (*9*), and Waymo's self-driving car (*10*) may continue this trend and provide even greater comfort and safety benefits (*11*, *12*). The safety impact (i.e. the reduction in crashes resulting in injuries or deaths) of these automated vehicle technologies will be limited by the ability of the automation and human to perform their responsibilities, their ability to transfer control to and from one another, their ability to conduct appropriate interactions with the transportation network, and appropriate levels of trust and reliance of the human driver on the automation (*13*–*18*). The human factors and driving safety research communities have primarily focused on transitions of control (*15*, *19*) although some efforts have been made to analyze interactions between vehicles in automated mode and the transportation network (*17*, *20*). In parallel with these efforts, companies have continued to pursue the development and testing of automated vehicle technologies on public roadways. In the state of California, USA, these tests must be documented in two reports documenting automation disengagements and crashes (*21*). The disengagement reports document the number and nature of transitions of control for each company and the crash reports contain a set of required fields along with an unstructured narrative report. The databases containing these reports offer a unique opportunity to augment the findings from controlled laboratory and on road studies.

Prior research has used these databases to compare automated and manual vehicle safety (*11*, *22*, *23*), analyze transitions of control (*24*–*26*), and identify trends in collision types and crash dynamics (*26*). These studies have found that automated vehicles tend to be safer than manually driven vehicles, although there are open issues regarding the impact of transitions of control on safety (*23*). Analyses of the crash database have found that the majority of crashes occur when the vehicle in automated mode is stopped, that approximately 20 % of the crashes occur following a manual transition, and that no crashes have been reported during an automated initiated transition of control to the driver (*25*, *26*). The findings have been derived primarily from frequency analyses of the fields in the crash database and manual reviews of crash narratives. This focus is limited because the fields are predefined based on prior knowledge, and the inherent bias present in manual reading and coding. For example, weather, lighting, and road conditions were not added as a field until 2018, although they are mentioned in several crash narratives prior to their inclusion.

One method of overcoming these limitations is topic modeling analysis (*27*). Topic modeling is a set of automated algorithms that process text and identify common topics or themes across a dataset. These themes, in turn, can be used to make inferences about the database contents that extend beyond pre-defined fields. In the transportation domain, topic modeling analyses have been applied to vehicle owners complaints (*28*), naturalistic driving data (*29*), and consumer survey responses to automated vehicles (*30*). These analyses have effectively identified themes consistent with notable incidents (e.g., the Ford Firestone tire recalls), driving groups, and core consumer concerns. The promising results of prior applications of topic modeling in the transportation domain suggest that it may be effective for identifying themes in automated vehicle databases. The goal of this study is to assess this suggestion by conducting a topic modeling analysis of the California Department of Motor Vehicles (CDMV) automated vehicle crash report narratives. The goals of this analysis are to highlight the benefits and limitations of topic modeling with the current data in the database, offer insights into future research and development needs, and provide a guide for future analyses of the database.

# 2 MATERIAL AND METHODS

## 2.1 California Department of Motor Vehicles Database

The CDMV vehicle disengagement reports and crash report database are stored in a publicly available web repository (*31*). While the disengagement reports are informative, they do not contain unstructured narrative descriptions and are thus not amenable to topic modeling analyses. Given this limitation, the analyses reported here focus on the crash reports. The crash reports contain the following details:

1. Manufacturer, make, model and year of the automated vehicle as well as any other vehicles involved in the collision
2. City, street, and intersections of the incident location
3. Movement of automated vehicle and the other vehicles involved prior to the collision
4. Date and time of the crash
5. Weather, lighting, and roadway conditions at the time of the collision (available after April 2018)
6. Type of the object involved in the collision (e.g., vehicle, cyclist, pedestrian)
7. Injuries and property damages
8. Crash narrative describing the circumstances of the incident

The analysis reported here considers all the reports from October 2014 to June 2019. This dataset includes 167 automated vehicle crashes, all of which occurred in the San Francisco Bay Area (primarily in San Francisco and Mountain View) in both automated and conventional (i.e. manual) mode. The current study focuses on situations where the crash occurred while the focal vehicle was in automated mode or during the transition of control from automated to conventional mode. Crashes in automated mode were identified based on the corresponding field in the database. Crashes during transitions were identified through manual reviews of the narratives. For all crashes a consensus on classification was achieved via discussions between the authors. Crashes where the focal vehicle was in manual mode were excluded because of the current focus on connecting the crash narrative trends to studies of transitions and automated driving. Furthermore, nearly all of the crashes in manual mode occurred while the focal vehicle was stopped and thus are unlikely to be informative for safety. The elimination of these crashes reduced the dataset to 114 crash reports (18 transition crashes, 96 crashes in automated mode).

**2.2 Probabilistic topic modeling analysis**
The crash narratives of each report were analyzed through a probabilistic topic modeling (PTM) approach. Probabilistic topic modeling is an unsupervised method of identifying themes in a collection of narratives (*32*). PTM models can be characterized by a set of distributions over terms, topics, and narratives. Fitting a PTM model consists of learning the probability densities for the narratives and terms over the topics. In this way, narratives and terms can be associated with multiple topics. Graphical depictions of the topic-term and topic-narrative distributions can be used for model inference. PTM is an unsupervised learning technique and thus it provides one plausible view of the dataset, rather than an optimal view. PTM has been successfully applied to journal articles (*32*), and textbooks (*33*). In these applications, success is typically defined according to the model's ability to identify coherent topics and provide substantial insight to the dataset. The current article uses PTM to identify common elements of automated vehicle crash narratives (e.g., the maneuvers that led to the crash) that may provide additional perspective on the other crash characteristics in the dataset. The PTM approach in this study used the "tm" (*34*, *35*) and "topicmodels" (*36*) packages in R 3.6.1 (*37*). The steps of this process include:

1. Transcribe the crash narratives from the database into a comma separated value format.
2. Convert the comma separated value dataset into a corpus (i.e. an R object representing a collection of documents).
3. Remove all punctuation in the corpus.
4. Convert the words in the corpus to lower case.
5. Remove *stop words*.
6. Convert the corpus to a term-document matrix.
7. Select the number of topics.
8. Process the data with Latent Dirichlet Allocation (LDA) to fit a model.

Steps 1-5 were completed with the "tm" package (*34*, *35*) function "tm_map". Steps 1-4 are standard practice in natural language processing analyses. Step 5—*remove stop words*—involves removing words from the dataset directly. The goal of this removal is to eliminate words that are not informative for crash categorization. In other natural language processing analyses (e.g., Blei et al., 2003) these stop words consist of common English words (e.g., "the", "and", "into"). In this analysis additional terms were added to this list that were not focused on the characteristics of the crash. These words corresponded to events regarding the parties involved in the crash (e.g., "autonomous," "vehicle") or the reporting procedure (e.g., "exchanged," "insurance," "crashed"). Words that specifically referred to company names or vehicle models were also removed from the data. An example set of stop words is provided in Table 1. The remaining steps were completed with the "topicmodels" package (*36*) in R. Step 6 consisted of converting the corpus into a term document matrix, which maps the frequency of distinct words occurring in each crash narrative. Following the term document matrix creation, the number of topics were selected.

Table 1 List of stop words and the rationale for their exclusion

| Type | Stop words | Rationale |
| --- | --- | --- |
| Common English words | and, the, in, on, should, do, while, just, either, other, under, when, each | Common English words that occurred at a high frequency and are likely to be uninformative (using the "English" set from the R stopwords package). |
| Crash descriptive | damage, involved, driver, mph, travelling, accident, collision, contact, sustained | Terms regarding vehicles involved in the collision, occurred at a high frequency across all narratives. |
| Automated vehicle descriptive | autonomous, mode, operating, AV, vehicle, model, technology | Nearly all of the reports contained a sentence describing the AV involved along with the mode. Thus, it is unlikely these terms would differentiate narratives. |
| Common police involvement descriptions | police, called, report, exchanged, experienced, information, parties, injuries | Most reports discussed the extent of police involvement and insurance information exchanges. Thus, these descriptions were unlikely to differentiate across narratives. |
| Vehicle make and model | Google, Waymo, Zoox, Lexus, Dodge, Volvo, Toyota, Chrysler, Ford, Subaru, Honda, Mercedes, Cruise | Manufacturer name or vehicle make were explicitly not considered in this analysis, as there are insufficient samples across manufacturers to make substantial conclusions about their relative safety. |
| Locations and street names | California, Oregon, Mountain View, San Francisco, San Antonio, El Camino Real, Rengstorff, Shoreline, Bryant, 3$^{rd}$, 10$^{th}$ | The crash locations were either common across most of the database or only occurred in a single report, thus they were unlikely to be informative for this analysis. |

The number of topics is a model input and is typically determined through a priori understanding of the dataset or a sensitivity analysis (*32*). In this analysis, the optimal number of topics was selected through a hybrid model of sensitivity analysis and topic overlap. The sensitivity analysis consisted of an iterative investigation of model fit statistics by the number of topics in the model. The model fit was assessed using four metrics widely used in the natural language processing and topic modeling literatures (*39–42*). Figure 1 shows the results of this sensitivity analysis, depicting the normalized values of the fit metrics by the number of topics in the model. In all cases the minimum value of the fit function represents the model with the best fit; however, each additional topic decreases the degrees of freedom in the model. Similar to other statistical analyses, the goal of topic number selection is to maximize fit with the fewest number of topics. While there is significant variance in the number of topics across the metrics, individual methods suggest an optimal number of topics of 2, 5, 10, or 13. Of these candidate topics, a model with 5 topics was selected because the 10 and 13 topic models included several topics with the same most strongly associated word, and the two topics model provided little insight to the data.

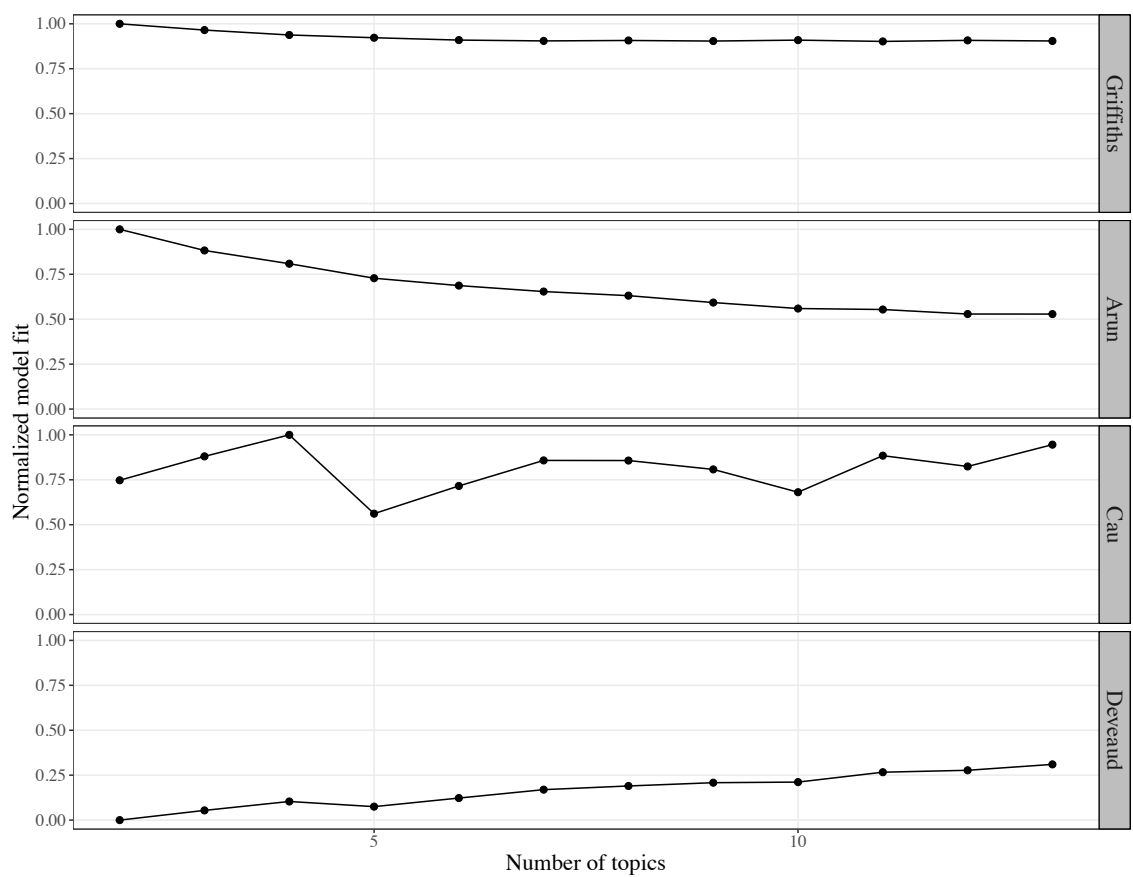

Figure 1 Number of topics and corresponding normalized model fit

## 2.3 Model Validation

The validity of the topic model fit was assessed through two methods: manual analysis of the narratives associated with each topic and comparisons with the fixed fields of the database. These analyses were captured by frequency counts. For example, if a topic was associated with the terms "rear," "end," and "crashes," the narratives were reviewed to identify the percentage of rear-end crashes associated with the topic. While such analysis does not provide formal validation of the model, it does lend credence to the robustness and viability of the conclusions derived from the model. Further, it provides an indication of the model's ability to replicate a manually intensive review of the narratives.

## 3   RESULTS

The results of the topic modeling analysis are summarized in Figure 2, Figure 3, and Table 2. Figure 2 shows the five topics (one graph per topic) characterized by the 10 words most strongly associated with each topic. The terms "manual," "control," and "took" in topic 1 suggests that the topic primarily captures crashes in transitions. A manual analysis of these crashes showed that 50% of crashes associated with this topic are manual disengagement crashes. The types of crashes were distributed across right-side (35%) and left-turn (20%) collisions at intersections (38%). Topics 2 and 3 are both represented by the terms "bumper," "rear," "rear ended," "intersection," and "light". These terms indicate that these topics contain crashes in which an automated vehicle was struck from the rear when stopped at a traffic signal at an intersection (86% of crashes associated with topic 2 and 80% of topic 3 crashes were rear-end collisions). The distinction between the topics is "right turn" which appears in topic 2 but not topic 3, indicating the topic 2 is more strongly associated with rear-end crashes in the right turn lane. The terms "left," "side," and "overtake" in topic 4 depict a side swipe collision from the left side while overtaking. The crash frequencies show that 56% of crashes in this topic are side swipe and cut-in collisions. The term "motorcycle" in this topic indicates that a motorcycle is one of the vehicles involved in the crash (20% of topic 4 crashes). Topic 5 captures the crashes that happened at an intersection (50% of topic 5 crashes). The topic is distinct from topics 2 and 3 because of the inclusion of oncoming traffic, which was present in two scenarios. The first scenario involved the vehicle in automated mode yielding to oncoming traffic at an intersection and being struck from the rear by another vehicle. The second scenario, which occurred in two crashes, consisted of a vehicle colliding with the vehicle in automated mode while it tried to move into the oncoming lane to overtake the stopped vehicle in automated mode. Table 2 further illustrates these topics by providing a sample of crash narrative correlated to each topic, samples are categorized by the automated vehicle mode and motion at the time of the crash. The dates of the crashes are provided as a key for the original crash report links.

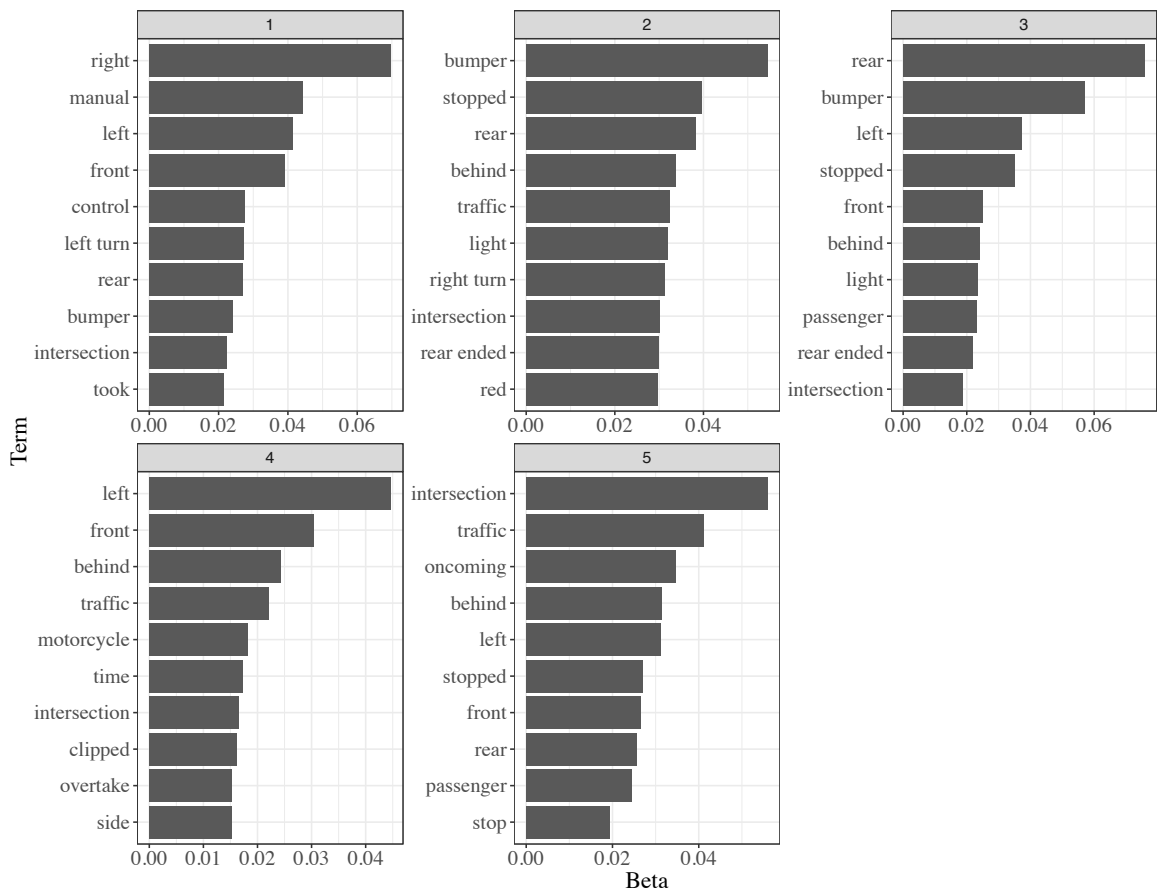

Figure 2 Distribution of terms across the five topics

Table 2 Crash narrative samples per topic. Note. AV stands for autonomous vehicle. All the locations, manufacturer, make, and model of the vehicles are removed from the table. The complete crash narratives can be accessed through the CDMV database (*43*)

| Topic | Mode | Motion | Crash date | Crash narrative |
|---|---|---|---|---|
| 1 | Transition | Moving | 14 Sep 2016 | "… As [the AV] was completing a lane change in autonomous mode from the far right lane to the middle lane … near [an] intersection, a car stopped in traffic in the far left lane … abruptly changed lanes into the middle lane, immediately in front of [the AV]. [The AV] test driver took manual control of [the AV] and quickly merged back into the far right lane to avoid the vehicle. Another vehicle approaching from behind in the right lane … then struck the rear passenger side quarter panel of [the AV]. The other vehicle sustained moderate damage to its front bumper ..." |
| 2 | Autonomous | Moving | 7 April 2015 | "[An AV] was traveling … in the rightmost lane … and came to a complete stop for a red light at [an] intersection …. [The AV] then proceeded to make a right turn on red by creeping forward to obtain a better field of view of cross traffic …. While creeping forward, [the AV] detected a vehicle approaching … and came to a stop in order to yield to the approaching vehicle. [The AV] was just starting to move when the vehicle following immediately behind it, which was also attempting to make a right turn …, failed to brake sufficiently and struck [the AV]'s bumper…" |
| 3 | Autonomous | Stopped | 24 Sep 2018 | "[An AV] was stopped … at a traffic light at [an] intersection. when a box truck made contact with the rear bumper of [the AV]. [The AV] sustained minor damage to the rear bumper, and the box truck had minor damage to the front bumper …" |
| 4 | Autonomous | Moving | 7 Dec 2017 | "[An AV] in heavy traffic, was involved in a collision while just past [an] intersection. [The AV] was traveling in the center of three one-way lanes. Identifying a space between two vehicles (a minivan in front and a sedan behind) in the left lane, [the AV] began to merge into that lane. At the same time, the minivan decelerated. Sensing that its gap was closing, [the AV] stopped its lane change and returned fully to the center lane. As [the AV] was re-centering itself in the lane, a motorcycle that had just lane-split between two vehicles in the center and right lanes moved into the center lane, glanced the side of [the AV]. The motorcyclist was determined to be at fault for attempting to overtake and pass another vehicle on the right…" |
| 5 | Autonomous | Stopped | 15 Nov 2018 | "[An AV] was waiting for oncoming traffic to clear the intersection before making a left turn … when another vehicle made contact with the right rear corner of [the AV] while attempting to pass it, causing damage to [the AV's] rear bumper…" |

Beyond the associations between terms and topics, it is also informative to analyze the associations between crash narratives and topics. Figure 3 represents the distribution of the crash narratives (y-axis) by the five topics (x-axis), grouped by vehicle mode (i.e., autonomous or transition) and motion (i.e. moving or stopped) at the time of the crash. The groups were identified based on the associated fields in the reports and included to highlight the notable findings in Figure 2. In Figure 3, dot size represents the strength of association between a topic and the narrative, with larger dots indicating larger probability density. This figure makes some of the term-topic associations clearer, including the fact that nearly all (94%) of the transition crashes are related to the first topic. In contrast with topic 1, most of the crashes associated with topic 2 and 3 happened in autonomous mode. Topic 2, 5, and, to a lesser extent, topic 3 are mostly associated with collisions where the automated vehicle was stopped. Again, this finding supports the results of term-topic analysis in which rear-end collisions at intersections where one vehicle was stopped were captured by these three topics. The majority of crashes associated with topic 4 happened while the autonomous vehicle was in motion. This finding also aligns with the results of the term-topic associations and the theme of cut-in and side swipe collisions. Finally, Figure 3 shows that several crash narratives share topics. For instance, topic 1 and 4 are both associated with several crash narratives. This is notable as it illustrates that a substantial portion of the transition crashes are also associated with side-swipe collisions on overtake. Collectively these results show that in addition to rear-end collisions at intersections, crashes associated with manual transitions, and crashes involving a side-swipe during overtake are prominent themes in the database.

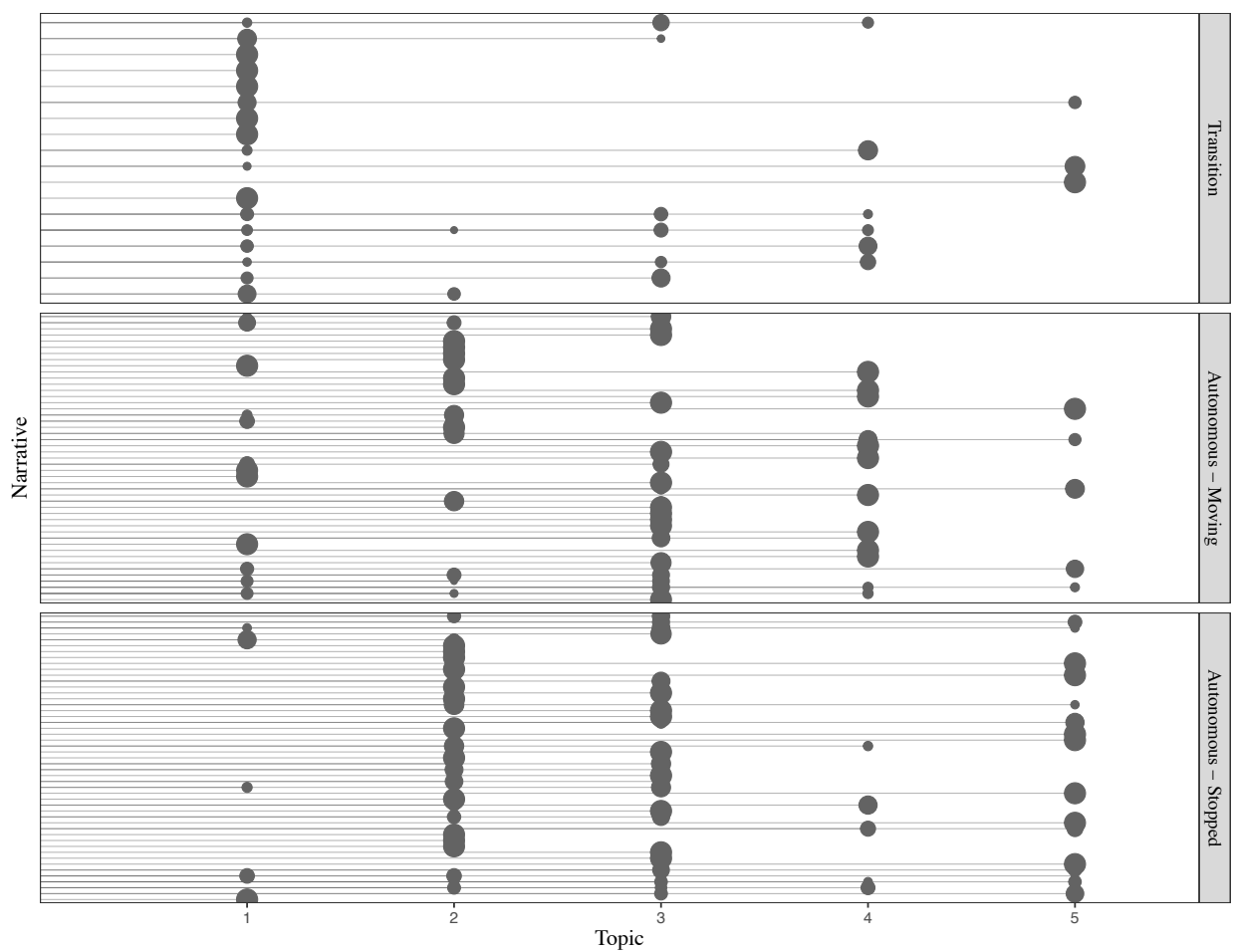

Figure 3 Distribution of topics per crash narratives. The dot size represents the strength of association (i.e. probability density on a topic for a given narrative) between the topic and the report, with larger dots representing stronger associations.

## 4 DISCUSSION

The primary goal of this study was to assess the utility of probabilistic topic modeling analysis of the California Department of Motor Vehicles automated vehicle crash database. The results suggested that the database was best described by 5 topics which were associated with: transitions of control, collisions at intersections in a right-turn lane, collisions at intersections in non-turn lanes, side-swipe crashes during a left-overtake, and collisions at intersections involving oncoming traffic. While the duplication of topics involving crashes at intersections may seem concerning, it is not unexpected given that previous studies have shown that the majority of the crashes in the database are rear-end collisions at intersections (*24*, *26*). The transition topic is important as it builds on preliminary frequency analyses that have found manually initiated transitions represent approximately 20% of the crashes in the database (*25*). This analysis suggests that the number of crashes associated with transitions now represent nearly 30% of the database. Further, the appearance of transitions as a topic suggests that these transitions may be a significant impediment to the crash risk and safety associated with automated vehicles. The side-swipe collision on overtaking topic is a novel finding of this analysis. Given that these collisions inherently involve interactions with the transportation system, this finding adds to the growing concerns associated with the safety of automated vehicles interacting with the existing transportation social network (*16*, *17*). Beyond the general concerns that arise from the emergence of this topic, it is specifically notable that motorcycles emerged as a strongly associated term. As the example of topic 4 in Table 2 indicates, not all of these crashes are severe, however their appearance warrants future attention and monitoring. It is further concerning that a number of crashes associated with the side-swipe topic are also associated with the transition topic. This finding may partially validate concerns over the extensive workload placed on drivers in automated mode tasked with monitoring challenging roadway conditions and an automation (*44–46*). The collective results suggest that topic modeling is a viable approach for analyzing the CDMV database and that it can lead to novel insights. The emergence of themes across topics and the combinations of crash locations (e.g., intersections), vehicle maneuvers (e.g., overtake), vehicles involved (e.g., motorcycles), and automation modes (e.g., transitions) involved in each topic illustrate the unique view that topic modeling can provide that extends basic frequency analysis.

The findings also have several implications for future research. Prior reviews of the literature suggest that the majority of studies on transitions of control have focused on automated-to-manual transitions initiated by the automation and preceded by an alert (*15*, *19*). The results of this analysis suggest that with current technology few crashes occur following these types of transitions. This analysis suggests that the majority of crashes during transitions in the database occurred during driver-initiated transitions rather than automation-initiated transitions. Thus, there is a significant need for additional research on manually initiated transitions. The association of the words "right," "left," and "intersection" within the transition topic (topic 4) suggest that future studies should investigate transitions during turns and at intersections. The side-swipe overtake topic also contrasts with the focus of prior studies. The majority of prior transition studies have focused on "obstacle reveal" scenarios in simple traffic conditions (15). The appearance of side-swipe overtake crashes in this analysis suggests a need for further studies investigating these situations. The occurrence of motorcycle crashes within this topic suggests a need to investigate the interactions between vehicles in automated mode and motorcycles. While interactions between motorcyclists and manually-driven vehicles have been studied (*47–49*), the authors are unaware of studies involving automated vehicles. The results here suggest that future studies should focus on interactions during side-swipe crashes and lane changes, particularly with lane-splitting motorcycles. It is also notable that the majority of crashes associated with this topic occurred when the vehicle was in automated mode and was moving. This further illustrates the need for investigations of silent automation failures in which the automation fails to detect an imminent crash and requires driver input.

Beyond the identification of core areas for future work, the topic modeling analysis discussed here provides a direct practical contribution as a tool for continuously monitoring safety issues with automated vehicles similar to the proposal of Ghazizadeh et al. (*28*). The automated nature of this tool may allow it to track the evolution of crash themes and, potentially, provide an early warning for risky road environments or common failures. With the current growing rate of crashes observed in the database (46 crashes/year), manually analyzing the reports will soon become infeasible. The results here suggest that probabilistic topic modeling can identify themes that are consistent with manual analysis. Thus, future work should investigate the use of probabilistic topic modeling for monitoring, particularly with regards to alternative topic modeling methods, parameter optimization, and the evolution of themes with larger datasets.

While this analysis highlighted important areas for further investigation, the results are limited by several factors. The total number of crashes reported in the database (167) is small relative to many other topic modeling analyses, which often include hundreds—or thousands—of narratives. This limited set of narratives may provide one explanation for the significant overlap between several topics. Another broad limitation of this analysis is that the database contains only crashes in California in vehicles driven by test drivers, and may not be representative of crashes in other locations or with the general population. In addition to the limitations of the database, several steps of the probabilistic topic modeling approach require subjective input. We attempted to mitigate the impact of these choices through a sensitivity analysis for topic number selection, and a careful stop words selection process; however, different choices in these processes would lead to changes in the results. As such, the topic modeling results here should be interpreted as one way to view the data, rather a definitive depiction. Future work may address these issues by incorporating other automated vehicle crash databases, conducting a cross-validation analysis of the topic modeling results, and investigating alternative topic modeling approaches such as structural topic modeling (*27*).

## 5 CONCLUSIONS

In this article we analyzed automated vehicle crash narratives from the California Department of Motor Vehicles crash database, and compared the results to identify safety concerns and gaps in the current research. Topic modeling analysis identified themes of manual transitions, sideswipe crashes while being overtaken from the left side, and rear-end collisions while turning at signalized intersections. These findings suggest that future empirical studies should investigate driver-initiated transitions of control, overtaking scenarios, silent failures, complex traffic situations, and interactions with motorcycles. The success of topic modeling in identifying these topics suggests that it may be a useful tool for monitoring safety issues in the database automatically.

## 6 ACKNOWLEDGEMENTS

Support for this research was provided in part by a grant from the U.S. Department of Transportation, University Transportation Centers Program to the Safety through Disruption University Transportation Center (451453-19C36).